\documentclass[aps,prd,twocolumn,nofootinbib]{revtex4}
\usepackage{amsmath,amssymb,amsfonts,epsfig}
\usepackage{color,graphicx}
\usepackage{subfigure}
\usepackage{bm}
\usepackage{multirow}

\newcommand{\be}{\begin{equation}}
\newcommand{\ee}{\end{equation}}
\newcommand{\bea}{\begin{eqnarray}\displaystyle}
\newcommand{\eea}{\end{eqnarray}}

\newcommand{\bhat}[1]{\hat{\mathbf{#1}}}

\def\d{{\rm d}}

\def\LCDM{$\Lambda$CDM}

\begin{document}

\title{Orbifold Line Topology and the Cosmic Microwave Background}

\author{Ben Rathaus}
\email{ben.rathaus@gmail.com}
\author{Assaf Ben-David}
\email{bd.assaf@gmail.com}
\author{Nissan Itzhaki}
\email{nitzhaki@post.tau.ac.il}

\affiliation{Raymond and Beverly Sackler Faculty of Exact Sciences,
School of Physics and Astronomy, Tel-Aviv University, Ramat-Aviv, 69978,
Israel}

\begin{abstract}

We extend our study of a universe with a non-classical stringy topology, and consider an orbifold line topology, $\mathbb{R}\times\mathbb{R}^2/\mathbb{Z}_p$. This topology has a fixed line and identifies each point in space with $p-1$ other points. An observable imprint of an orbifold line on the CMB is the appearance of up to $(p-1)/2$ pairs of matching circles. Searching the WMAP data for matching circles, we can rule out an orbifold line topology with $p$ up to 10, except for $p=8$. While the significance of the peak at $p=8$ varies between data releases of WMAP, it does not appear in Planck data, enabling us to rule out $p=8$ as well.
\end{abstract}

\maketitle

\section{Introduction}

It has been known for quite some time \cite{Cornish:1997rp, Cornish:1997ab, Cornish:2003db, ShapiroKey:2006hm, Bielewicz:2010bh, Vaudrevange:2012da,Mota:2011hw} that matching circles in the cosmic microwave background (CMB) sky is a possible way to detect a non-trivial topology of the universe. So far, the main focus was on classical topologies that contain no singularities, such as $\mathbb{R}^2\times \rm{S}^1$ \cite{Starobinsky:1993yx, deOliveiraCosta:1995td, Cornish:1997rp, Park:1997ad, Cornish:1997ab, Cornish:2003db, Phillips:2004nc, ShapiroKey:2006hm, Aurich:2010wf, Bielewicz:2010bh, Vaudrevange:2012da,Riazuelo:2003ud, Mota:2011hw}. In string theory, however, the class of allowed topologies is larger, as string theory resolves a certain type of singularities, known as orbifolds. 

In a previous work \cite{BenDavid:2012kr} we have considered the possibility that the topology of the universe is that of an orbifold point, and found limits on the distance to the orbifold point. Here we extend our search for an orbifold line.

As was shown in \cite{BenDavid:2012kr}, while imposing an orbifold symmetry on the standard inflationary setting breaks both homogeneity and isotropy, it does not break Gaussianity. Such a symmetry only changes the two-point correlation function of the fields.

Furthermore, as a symmetry with fixed points is only consistent in string theory, one would expect stringy degrees of freedom to affect the inflaton potential. Indeed, the effects of stringy models of inflation have been extensively studied in various works, e.g. \cite{Baumann:2007ah,Hertzberg:2007ke,McAllister:2007bg,Silverstein:2008sg,Baumann:2009ni,McAllister:2008hb,Burgess:2011fa} (for a very recent discussion on string inflation after Planck, see \cite{Burgess:2013sla}). However, as discussed in \cite{BenDavid:2012kr}, in the setting considered here, the extra degrees of freedom are confined to the fixed points and do not propagate away. As long as we assume that the string scale is much smaller than the Hubble scale during inflation, the confined stringy modes do not affect the inflaton power spectrum. This way, as long as we are not interested in the physics close to the fixed points, we can simply impose the orbifold symmetry on the existing fields in the usual inflationary setting without affecting the potential and regardless of the extra degrees of freedom.
We stress that the model explored in this work does not provide an alternative framework for the recreation of matter after inflation, nor does the discovery of a stringy topology falsify inflation in any way. The model merely predicts how an orbifold line topology will be manifested in the CMB when considered on top of the standard \LCDM{} cosmology in a generic inflationary setting (for more details see \cite{BenDavid:2012kr}).

The paper is organized as follows. In \S\ref{sec:orbiLine} we review the observable imprints of an orbifold line topology, and devise a score to efficiently search for these imprints. We summarize the results of testing this statistic on WMAP data in \S\ref{sec:results}, compared with results from Monte Carlo simulations. Finally, in \S\ref{sec:discussion}, we discuss the results.

\section{Orbifold Line Topology}
\label{sec:orbiLine}

\subsection{Identification and Matched Circles}
\label{sec:ident}

\begin{figure}
\centering
\includegraphics[width=0.95\linewidth]{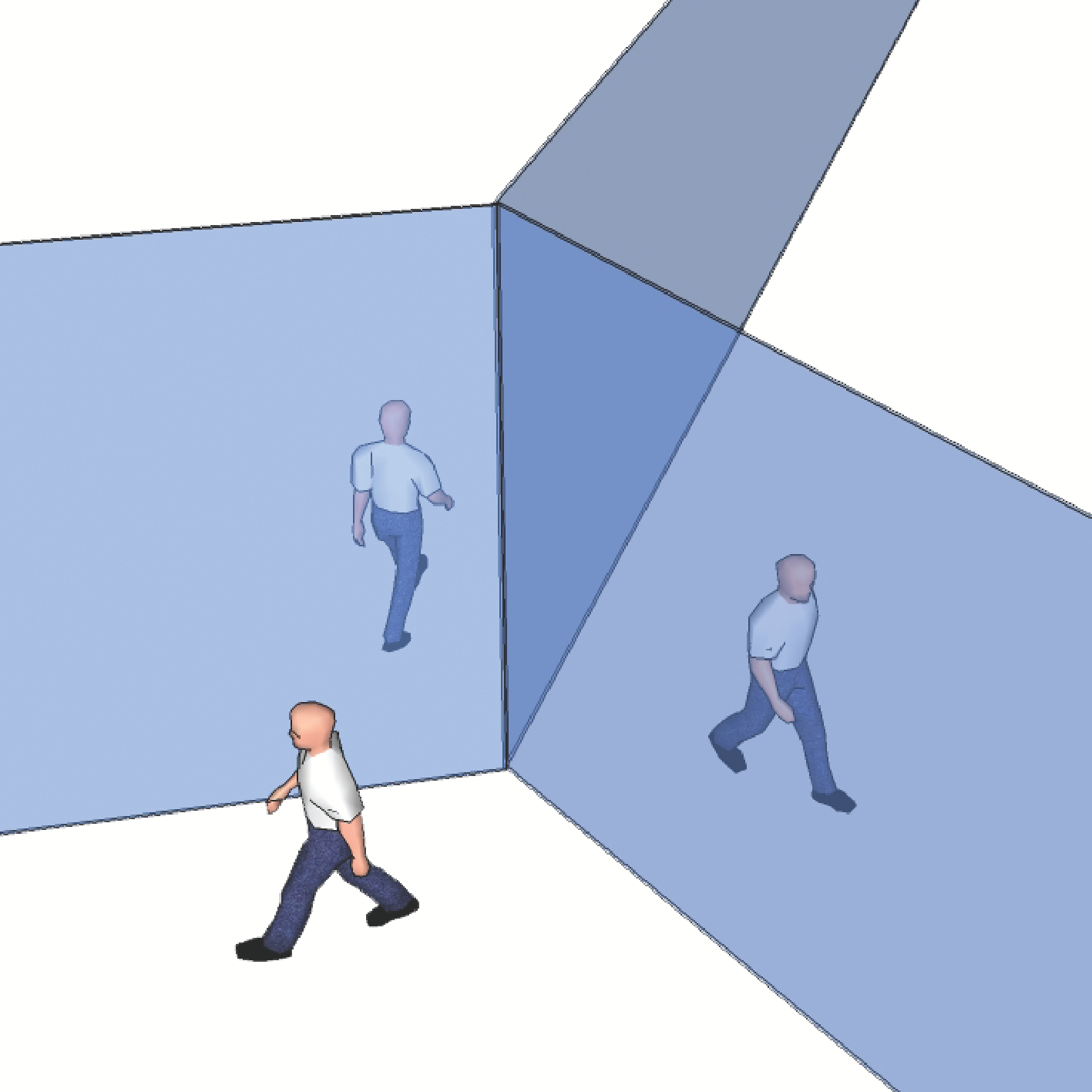}
\caption{An illustration of the 3 wedge-like replicated sectors and the fixed line of a $p=3$ orbifold line topology. Each point in space is identified with all other points that are rotated by an integer multiple of $2\pi/p$ around the fixed line, so the identification is between planes. \label{fig:simpleIllustration}}
\end{figure}

We consider an orbifold line topology, that can be expressed as $\mathbb{R}\times\mathbb{R}^2/\mathbb{Z}_p$, the quotient space under the action of the cyclic group $\mathbb{Z}_p$, where $p\ge2$ is an integer. This space is described by the metric
\be
\d s^2=\d z^2+\d r^2+r^2\d \varphi^2,
\ee 
with the identification
\be\label{eq:identPhi}
\varphi\sim\varphi+2\pi/p.
\ee
That is, $\mathbb{R}^3=\mathbb{R}\times\mathbb{R}^2$ is divided into $p$ replicated wedge-like sectors. Every point is identified with $p-1$ other points, one in each of the other sectors. This identification has a fixed line, where all the sectors meet (see Fig.~\ref{fig:simpleIllustration}).

\begin{figure*}
\centering
\includegraphics[width=0.45\linewidth]{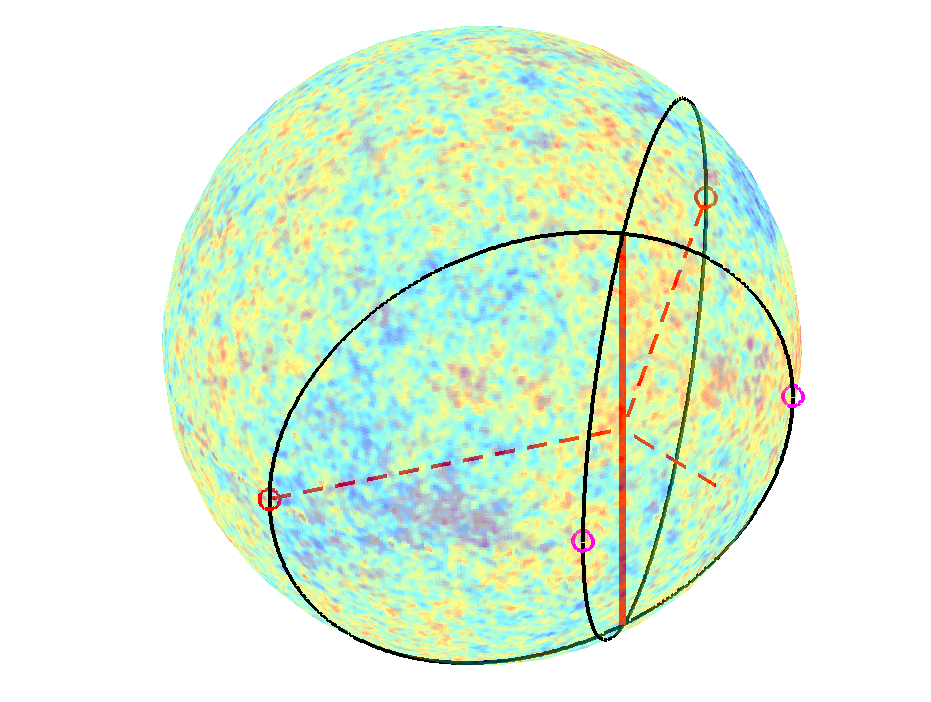}
\includegraphics[width=0.45\linewidth]{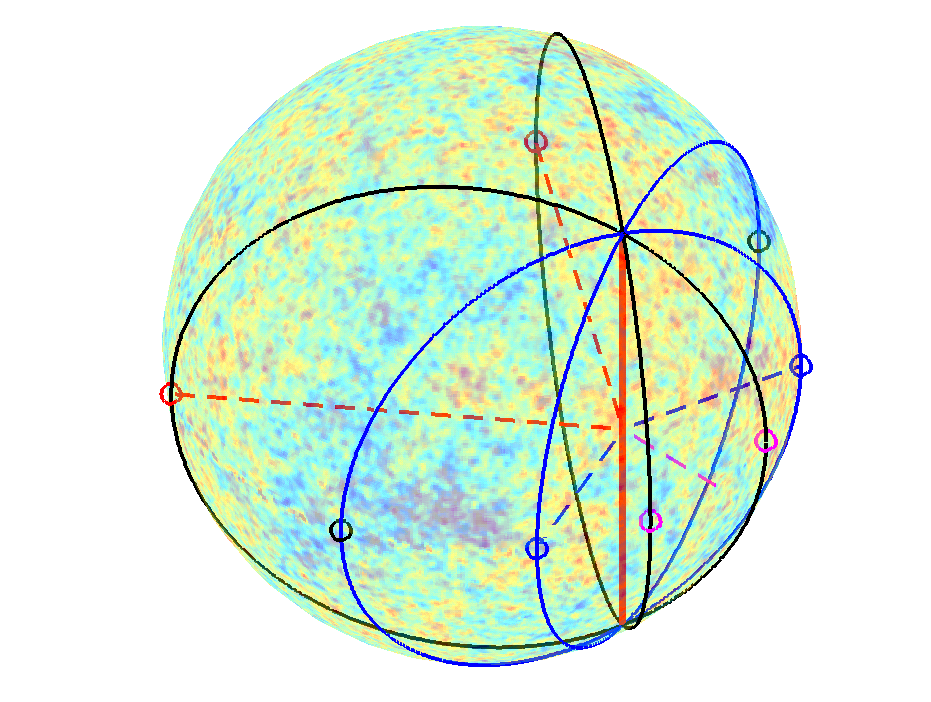}
\caption{An illustration of the matching circles in the CMB of an orbifold line at a distance of $r_0=0.6 r_*$, for $p=3$ ({\it left}) and $p=5$ ({\it right}). The fixed line is shown by a thick red line. The pair of black circles is the largest pair (largest separation and opening angles), and matching points along the circles are indicated in matching colors. For $p=5$ there is another pair of matching circles, shown in blue.\label{fig:illustration}}
\end{figure*}

It is apparent that the identification (\ref{eq:identPhi}) is between planes, and that the intersection of these planes with 
the last scattering surface (LSS) are circles. Therefore, if the universe has an orbifold line topology and
\be\label{eq:matchingThreshold}
r_0 < \frac{r_*}{\sin\pi/p},
\ee
where $r_0$ and $r_*$ are the distances to the line and the LSS, respectively,
we expect to find pairs of matching circles in the CMB sky, as illustrated in Fig.~\ref{fig:illustration}.
In fact, matching circles are still expected even if the line is outside the observable universe, $r_0 > r_*$, as long as (\ref{eq:matchingThreshold}) is satisfied. These circles, like any matching patterns caused by a non-trivial topology, have a characteristic width of $\sim 1^{\circ}$, corresponding to the thickness of the LSS. 

The separation angle between the centers of the two circles is $\pi(1-2/p)$. While for $p>2$ the two circles are distinct, for $p=2$ the separation angle vanishes, meaning that in this case, the two circles are actually the same one, viewed in opposite orientations. We consider such a circle as half a pair.

Furthermore, if $r_0 < r_*/\sin 2\pi/p$,  more than a single pair of matching circles will appear, but never more than $(p-1)/2$ pairs (with the definition of half a pair as above). The $n$-th pair will appear once $r_0 < r_*/\sin n\pi/p$, with a separation angle of
\be\label{eq:theta}
\theta_{p}^{(n)} = \pi(1 - 2n/p),
\ee
and with the opening angle $\alpha$ of each circle satisfying
\be\label{eq:alpha}
\cos\alpha = \frac{r_0}{r_*}\sin n\pi/p.
\ee

We note that for two integers $p>q\ge2$, such that $q$ is a divisor of $p$, two orbifold line topologies described by these integers will share {\it some} of the pairs of matching circles. The $n=1$ pair of the $p$-orbifold, however, is {\it never} shared with the $q$-orbifold. This pair has the largest separation and opening angles of all the $p$-orbifold pairs, and we refer to it as the largest pair.

\subsection{Score for Detecting Circles}

We test the match between the temperature profiles along two circles using the standard statistic, as in \cite{Vaudrevange:2012da,Bielewicz:2010bh,BenDavid:2012kr}. For each $p$, we focus on the largest pair of matching circles, $n=1$, and consider all line directions and orientations such that the circle-centers directions satisfy
\be\label{eq:n1n2Condition}
\bhat{n}_1\cdot \bhat{n}_2 = \cos \theta_{p}^{(1)}.
\ee
Since this pair has more pixels than other pairs, we expect it to yield a larger signal-to-noise ratio (S/N). In addition, we enumerate on all opening angles $25^{\circ} \le \alpha \leq 90^{\circ}$ in $0.25^{\circ}$ resolution. The distance to the line $r_0$ for each $\alpha$ is then determined by (\ref{eq:alpha}). Therefore, for given $p, \alpha$ we use the statistic
\be\label{eq:score}
S_p(\alpha) = \max\frac{2 \sum_{m} m T_{1,m}T_{2,m}} {\sum_{m}m\left( |T_{1,m}| ^2 + |T_{2,m}| ^2 \right)},
\ee
where $T_{i,m}$ is the Fourier transform of the temperature profile along the circle centered around $\bhat{n}_i$ with opening angle $\alpha$, according to $T_i(\phi) = \sum_m T_{i,m}e^{im\phi}$, and the maximization is over all $\bhat{n}_1,\bhat{n}_2$ satisfying (\ref{eq:n1n2Condition}). The relative phase between the circles is chosen such that $T_1(\phi=0)$ and $T_2(\phi=0)$ correspond to identified points.

The score $S_p(\alpha)$, however, is boosted for some portion of the $\alpha$-axis. As was noted in \cite{Vaudrevange:2012da}, this boost is a result of the intersection of the two circles due to the automatic match between segments of the circles. In the case of an orbifold line it occurs when $r_0 < r_*$. In fact, this effect is more prominent when the circles osculate rather than intersect, at $r_0\sim r_*$, even when the line is outside the LSS.

In order to address this issue we give less weight to parts of the circles that are close to each other. To do so we multiply the temperature profiles along the two circles by a weight function
\be \label{eq:window}
W_d(\gamma) = \frac{1}{2}\left[  \tanh (\gamma - d) + 1 \right],
\ee
where $\gamma$ is the angular distance between two corresponding points on the circles, and $d=2.5^{\circ}$. We apply this weight function before Fourier transforming the temperature profiles.

\section{Results}
\label{sec:results}

For the analysis in this paper we use the WMAP temperature maps \cite{wmap7, wmap9}. Our primary analysis is carried out on a map that consists of the 7-year foreground reduced W-band map outside the galactic mask Kp12 (covering $\sim 6\%$ of the sky), and the 7-year internal linear combination (ILC) map inside the mask. In addition, we compare some of the results with the 3, 5 and 9-year data, as explained below.

As mentioned in \S\ref{sec:ident}, the circles have a characteristic width of $\sim1^{\circ}$. Therefore, in order to maximize the S/N for detecting a pair of matching circles, we smooth the combined W and ILC map with a Gaussian beam of $1^{\circ}$ FWHM. The maximization over the directions $\bhat{n}_1, \bhat{n}_2$ is performed using a HEALPix \cite{healpix} $N_{\rm{side}} = 128$ search grid.

\begin{figure*}
\centering
\includegraphics[width=0.95\linewidth]{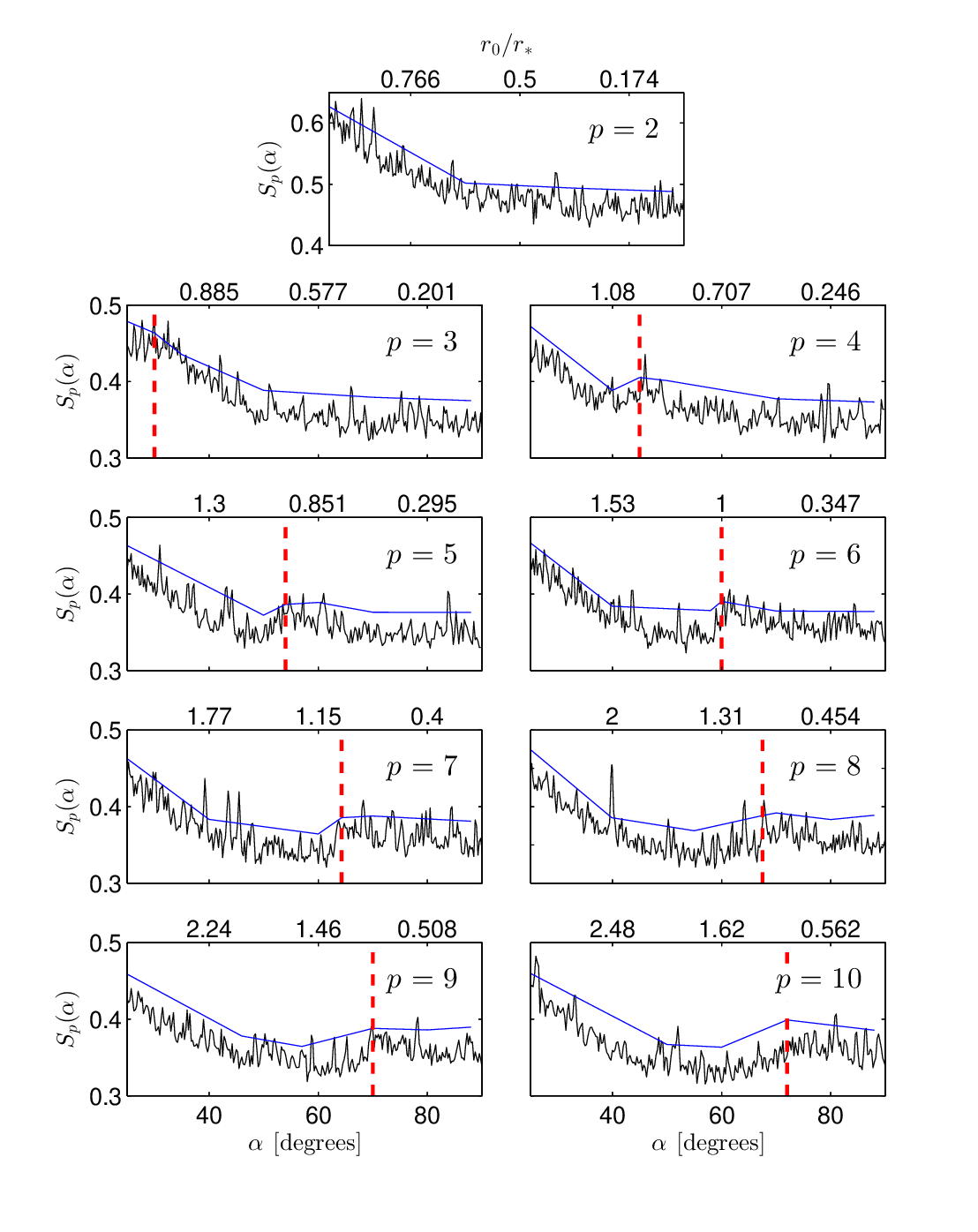}
\caption{The score $S_p(\alpha)$ for $p=2,\dots,10$ and for $25^\circ \le\alpha \le 90^\circ$. The vertical dashed red line designates $r_0=r_*$ for each $p$, except for the $p=2$ case, in which $r_0=r_*$ corresponds to $\alpha=0$. The top $x$-axis is the distance to the line $r_0/r_*$, related to the opening angle $\alpha$ (bottom $x$-axis) by (\ref{eq:alpha}). The blue line is the $95\%$ CL. The number of peaks crossing the $95\%$ CL is not anomalous, as we expect $5\%$ of all the $S_p(\alpha)$ evaluations to cross it. $S_8(39.9^\circ)$, however, is high and worth further investigation. \label{fig:allScores}}
\end{figure*}

The score $S_p(\alpha)$ for the 7-year data is shown in Fig.~\ref{fig:allScores} for $p=2,\dots, 10$, with a $95\%$ confidence level (CL) line, calculated using Monte-Carlo simulations.
We see that 31 peaks cross the CL line. However, we expect $5\%$ of all independent evaluations of $S_p(\alpha)$ to cross the $95\%$ CL line. Since our map is smoothed to a $1^\circ$ scale and we test 9 values of $p$, the number of random peaks expected to cross the CL line for the range $25^\circ \le \alpha \le 90^\circ$ is $\sim 65\times 9 \times 0.05 \simeq 30$. Thus the number of peaks crossing the line is not anomalous.

\begin{figure}
\centering
\includegraphics[width=0.95\linewidth]{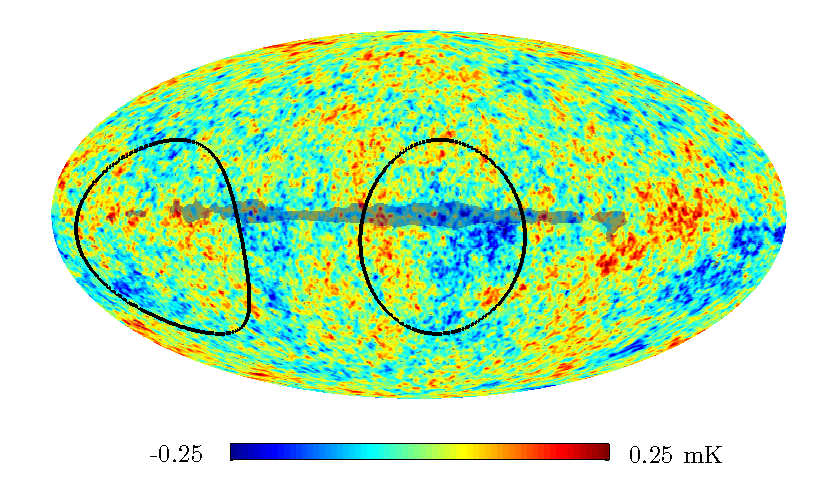}
\caption{The two circles that correspond to the high peak in the $p=8$ plot. The  temperature map in the background is the foreground reduced W-band map outside the galactic mask Kp12 (shaded area) and the ILC map inside the mask. The map is smoothed to $1^\circ$.\label{fig:p8Circles}}
\end{figure}

However, one peak is very high. The peak in the $p=8$ plot at $\alpha = 39.9^{\circ}$ is distinct, yielding $S_8(39.9^{\circ})=0.455$, and worth investigating. This peak is caused by a pair of circles centered around $(l_1,b_1) = (128^\circ,-10^\circ)$ and $(l_2,b_2) = (349^\circ,-10^\circ)$, as shown in Fig.~\ref{fig:p8Circles}, 
and corresponds to an orbifold line located in direction $(l,b) = (58^\circ,-27^\circ)$, at a distance $r_0=2.005 r_*$ away from us, outside the observable universe.

As previously stated, this peak might also be attributed to an orbifold line topology with $p'=8m$ where $m>1$ is an integer, in which case this pair of circles would not be the largest pair. We therefore check the largest pairs of $m=2,\dots ,10$ with the direction and distance to the line as above, and $\alpha$ according to (\ref{eq:alpha}). In all cases, $S_{p'}(\alpha) < 0.08$, meaning that this peak can only be attributed to a $p=8$ orbifold line. Additionally, we note that for $p=8$ there will be a second pair only if $r_0<\sqrt{2}r_*$, which is clearly not satisfied for this peak.

\begin{figure}
\centering
\includegraphics[width=0.95\linewidth]{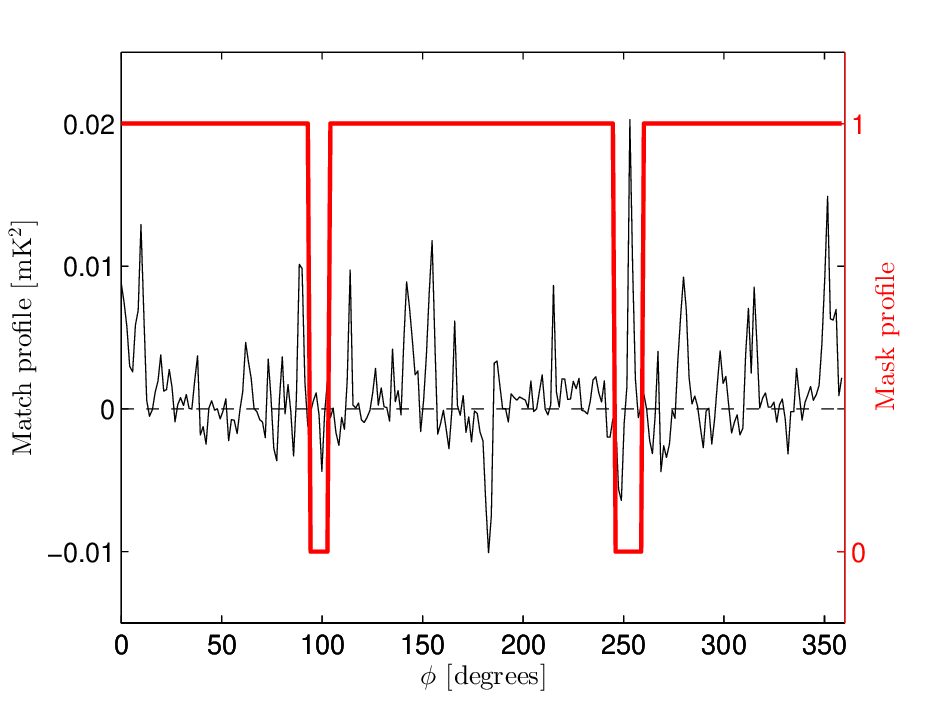}
\caption{The temperature match profile $T_1(\phi) T_2(-\phi)$ ({\it black}) and the mask values product profile ({\it red}) of the circle pair associated with the high peak in $S_8(\alpha)$. It can be seen that there is no single patch that dominates the match, but rather many contributions of similar amplitudes. \label{fig:ttProfile}}
\end{figure}

Before we estimate the significance of the peak, we first examine whether the match between the circles is due to some specific patches in them. Such a behavior would be inconsistent with the orbifold line model. For this sake, we plot in Fig.~\ref{fig:ttProfile}
the match profile $T_1(\phi) T_2(-\phi)$ as a function of the phase $\phi$ along the circles. We see that the product of the two temperature profiles is not dominated by a single patch, rather it consists of many contributions of similar amplitudes. In addition we superimpose the product of the Kp12 mask profiles along the same circles, smoothed to $3^\circ$. While the strongest contribution of the match profile is inside the mask, it does not contribute crucially to the score. Indeed, if we zero the temperature of the pixels responsible for this contribution and recalculate the score, we find that it changes  in only $2.5\%$. We conclude that the match profile test does not rule out the possibility that this candidate match is due to an orbifold line topology.

In order to evaluate the significance of the peak we compare it to random realizations. We simulate random maps taking into account the noise properties of WMAP. We generate an initial fluctuations map from the WMAP 7-year best fit power spectrum in the multipole range $2\le\ell\le1024$, at  HEALPix resolution $N_\text{side}=512$. We then smooth it with a Gaussian beam of $0.22^\circ$ FWHM, corresponding to the resolution of the W-band. Next, we add Gaussian noise, uncorrelated between pixels, with the same variance as that of the WMAP  data. Finally, we smooth the map to a $1^\circ$ resolution to match our data map.

We generate $90,\!000$ random maps and calculate $S_8(39.9^\circ)$ for each map. We  find that only one simulation gets a higher score than the WMAP7 data. This corresponds to a $4.4\sigma$ anomaly. To take the ``look elsewhere'' effect into account, we multiply this $p$-value by the number of independent evaluations in our search grid and get $10^{-5} \times 65\times 9 \sim2.75\sigma$. We can therefore say that for the 7-year data, the best candidate for a matching pair is marginally significant.

\begin{figure}
\centering
\includegraphics[width=0.95\linewidth]{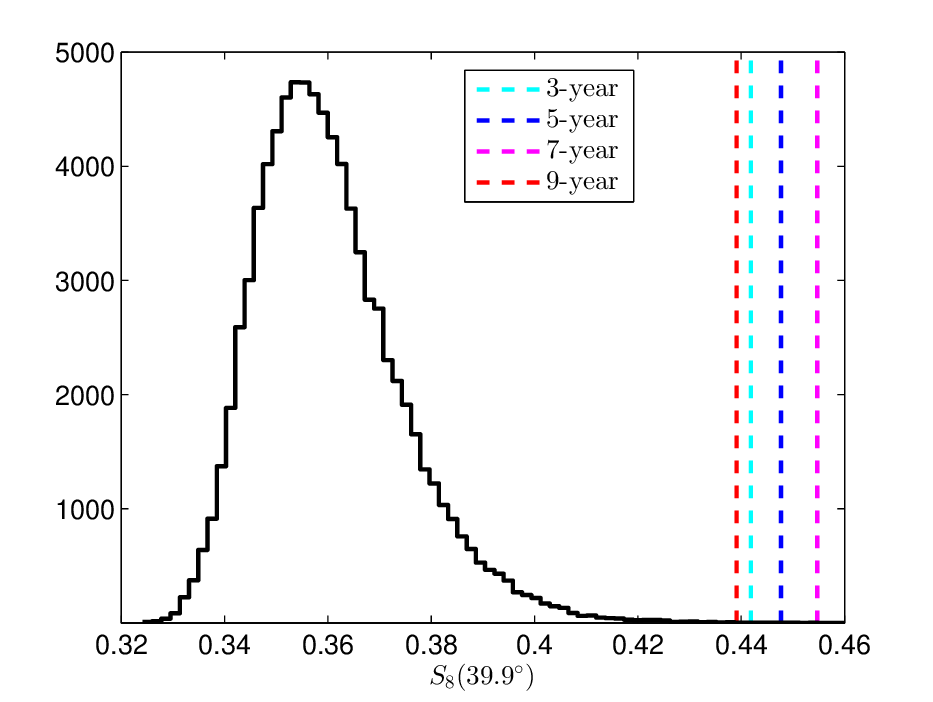}
\caption{The score $S_8(39.9^\circ)$ for the 3, 5, 7 and 9-year data, along with a histogram of the random simulations. \label{fig:histogram}}
\end{figure}

We would like to check whether this candidate appears also in the $3,5$ and recently published 9-year data of WMAP. We calculate $S_8(39.9^\circ)$ for the combined W+ILC map of each data release. Compared with the $0.455$ score of the 7-year data, we get $0.442, 0.448$ and $0.439$ for the 3, 5 and 9-year data, respectively. We plot the results, along with the histogram of the random simulations, in Fig.~\ref{fig:histogram}.
Thus, in all data releases this peak is observed and is fairly high. However, while the score increases for years 3, 5 and 7, as expected since the noise decreases for longer observation times, the 9-year score is the lowest. This motivates us to investigate further the source for this unexpected drop.

With this aim in mind, we wish to trace the discrepancy and to check whether it originates from the W-band map, or the ILC substitution inside the Kp12 mask. We do so by testing the score of the peak using the 9-year W-band map with the 7-year ILC inside the Kp12 mask, and vice versa. The results, summarized in Table \ref{tab:theTable},
\begin{table}
\centering
\renewcommand{\arraystretch}{1.5}
\renewcommand{\tabcolsep}{0.2cm}
\begin{tabular}{|l||c|c|c|}
\hline
 & ILC7 & ILC9 & Zeros\\
\hline
\hline
\multirow{2}{*}{W7} &   0.455 & 0.442 &  0.430\\
& ($4.39\sigma$)& ($3.86\sigma$)&($3.48\sigma$)\\
\hline
\multirow{2}{*}{W9} & 0.452 & 0.439& 0.427\\
&($4.23\sigma$)& ($3.82\sigma$) &($3.38\sigma$)\\
\hline
\end{tabular}
\caption{A summary of $S_8(39.9^\circ)$ for a few combinations of the 7 and 9-year WMAP data. The row designates the map used outside the Kp12 mask, and the column states the data filling the mask.
\label{tab:theTable}}
\end{table}
indicate that when we simply replace the data inside the mask from the 7-year ILC to the 9-year ILC, the score already drops significantly, and almost reaches its W9+ILC9 value. On the other hand, when testing the W9 map with the ILC7 inside the mask, the score drops by less than $0.5\%$ with respect to its W7+ILC7 value.

These results imply that the source of the drop in the score between the 7 and 9-year datasets may be the 9-year ILC map and not the W-band map. Looking at the ILC9 map  askance, we test if the strong signal is still found in the W-band map alone. Instead of using the ILC map, we now replace the data inside the Kp12 mask with zeros before evaluating the score. The results, also shown in Table~\ref{tab:theTable}, indicate that the contribution to the score from outside the mask is nearly unchanged ($0.7\%$) between the 7 and 9-year data. This supports our conclusion that the bulk of the signal is in the W-band map, and does not originate from the less reliable galactic plane area, while the drop in the score is mainly due to the ILC9 map.

Finally, as this work was being reviewed, the data from the Planck surveyor has become publicly available \cite{PlanckMaps}. While this new and extensive dataset deserves a complete analysis on its own right, we have felt it necessary to check the candidate $S_8(39.9^\circ)$ on the SMICA map of Planck as well. Using the SMICA map, smoothed to $1^\circ$ resolution, we get a score of 0.37, rendering this candidate insignificant. In this dataset, the peak that appeared in the $p=8$ plot for WMAP 7-year data does not show up.

\section{Discussion}
\label{sec:discussion}
In this work we have extended our previous search for a stringy topology \cite{BenDavid:2012kr} and considered an orbifold line. We have used the standard statistic (\ref{eq:score}) from e.g. \cite{Vaudrevange:2012da} to test the WMAP data for matching circles, as evidence for such a topology. Expecting the characteristic width of the circles to coincide with the thickness of the LSS, we have smoothed the data to a $1^\circ$ scale in order to maximize the S/N for detection.

Testing for topologies with $p=2,\dots,10$, we have found a single candidate in the 7-year data, with $p=8$, in direction $(l,b) = (58^\circ,-27^\circ)$ and at a distance $2.005r_*$ away. This candidate appears to be locally significant, at the $4.4\sigma$ level,  and taking into account the ``look elsewhere'' effect only marginally significant at the $2.75\sigma$ level. In other data releases of WMAP this candidate gets a high score as well. However, while for the 3, 5 and 7-year releases the score increases, the 9-year score is lower. We attribute this drop in the score to the data coming from the ILC9 map, from inside the galactic Kp12 mask, and not to the W-band data, outside the mask.

While the tension between these results and the assumption of a trivial topology is not high, the score for the candidate match found in the WMAP data does not allow us to rule out an orbifold line topology with $p=8$ using this dataset alone. However, testing the SMICA map of Planck shows no anomalous peak with $p=8$. Therefore, inclusion of the Planck dataset allow us to rule out all of the $p=2,\dots,10$ parameter space, constraining the allowed parameters for the model of an orbifold line topology to beyond this range.

\acknowledgments
We thank the anonymous referee for useful suggestions which helped improve the clarity of this paper. We also thank E.~D.~Kovetz for useful discussions. We acknowledge the use of the Legacy Archive for Microwave Background Data Analysis (LAMBDA) \cite{lambda}, as well as the Planck Legacy Archive (PLA) \cite{PLA}. This work is supported in part by the European Research Council (grant number 203247).


\begin{thebibliography}{28}


\bibitem{Cornish:1997rp}
  N.~J.~Cornish, D.~N.~Spergel and G.~D.~Starkman,
  Proc.\ Nat.\ Acad.\ Sci.\  {\bf 95}, 82 (1998)

\bibitem{Cornish:1997ab}
  N.~J.~Cornish, D.~N.~Spergel and G.~D.~Starkman,
  Class.\ Quant.\ Grav.\  {\bf 15}, 2657 (1998)

\bibitem{Cornish:2003db}
  N.~J.~Cornish, D.~N.~Spergel, G.~D.~Starkman and E.~Komatsu,
  Phys.\ Rev.\ Lett.\  {\bf 92}, 201302 (2004)

\bibitem{ShapiroKey:2006hm}
  J.~Shapiro Key, N.~J.~Cornish, D.~N.~Spergel and G.~D.~Starkman,
  Phys.\ Rev.\ D {\bf 75}, 084034 (2007)


\bibitem{Bielewicz:2010bh}
  P.~Bielewicz and A.~J.~Banday,
  Mon.\ Not.\ Roy.\ Astron.\ Soc.\  {\bf 412}, 2104 (2011)

\bibitem{Mota:2011hw} 
  B.~Mota, M.~J.~Reboucas and R.~Tavakol,
  Phys.\ Rev.\ D {\bf 84}, 083507 (2011)

\bibitem{Vaudrevange:2012da} 
  P.~M.~Vaudrevange, G.~D.~Starkman, N.~J.~Cornish and D.~N.~Spergel,
  Phys.\ Rev.\ D {\bf 86}, 083526 (2012)
  
\bibitem{Starobinsky:1993yx}
  A.~A.~Starobinsky,
  JETP Lett.\  {\bf 57}, 622 (1993)

\bibitem{deOliveiraCosta:1995td}
  A.~de Oliveira-Costa, G.~F.~Smoot and A.~A.~Starobinsky,
  Astrophys.\ J.\  {\bf 468}, 457 (1996)

\bibitem{Park:1997ad}
  C.~Park, W.~N.~Colley, J.~R.~Gott, III, B.~Ratra, D.~N.~Spergel and N.~Sugiyama,
  Astrophys.\ J.\  {\bf 506}, 473 (1998)

 
\bibitem{Riazuelo:2003ud}
  A.~Riazuelo, J.~Weeks, J.~-P.~Uzan, R.~Lehoucq and J.~-P.~Luminet,
  Phys.\ Rev.\ D {\bf 69}, 103518 (2004)
  
\bibitem{Phillips:2004nc}
  N.~G.~Phillips and A.~Kogut,
  Astrophys.\ J.\  {\bf 645}, 820 (2006)

\bibitem{Aurich:2010wf}
  R.~Aurich and S.~Lustig,
  Class.\ Quant.\ Grav.\  {\bf 28}, 085017 (2011)

\bibitem{BenDavid:2012kr} 
  A.~Ben-David, B.~Rathaus and N.~Itzhaki,
  JCAP {\bf 1211}, 020 (2012)



\bibitem{Baumann:2007ah} 
  D.~Baumann, A.~Dymarsky, I.~R.~Klebanov and L.~McAllister,
  JCAP {\bf 0801}, 024 (2008)
  [arXiv:0706.0360 [hep-th]].
  

\bibitem{Hertzberg:2007ke} 
  M.~P.~Hertzberg, M.~Tegmark, S.~Kachru, J.~Shelton and O.~Ozcan,
  Phys.\ Rev.\ D {\bf 76}, 103521 (2007)
  [arXiv:0709.0002 [astro-ph]].
 
  
\bibitem{McAllister:2007bg} 
  L.~McAllister and E.~Silverstein,
  Gen.\ Rel.\ Grav.\  {\bf 40}, 565 (2008)
  [arXiv:0710.2951 [hep-th]].
 
\bibitem{Silverstein:2008sg} 
  E.~Silverstein and A.~Westphal,
  Phys.\ Rev.\ D {\bf 78}, 106003 (2008)
  [arXiv:0803.3085 [hep-th]].
 
\bibitem{Baumann:2009ni} 
  D.~Baumann and L.~McAllister,
  Ann.\ Rev.\ Nucl.\ Part.\ Sci.\  {\bf 59}, 67 (2009)
  [arXiv:0901.0265 [hep-th]].


\bibitem{McAllister:2008hb} 
  L.~McAllister, E.~Silverstein and A.~Westphal,
  Phys.\ Rev.\ D {\bf 82}, 046003 (2010)
  [arXiv:0808.0706 [hep-th]].

\bibitem{Burgess:2011fa} 
  C.~P.~Burgess and L.~McAllister,
  Class.\ Quant.\ Grav.\  {\bf 28}, 204002 (2011)
  [arXiv:1108.2660 [hep-th]].

\bibitem{Burgess:2013sla} 
  C.~P.~Burgess, M.~Cicoli and F.~Quevedo,
  arXiv:1306.3512 [hep-th].


\bibitem{wmap7} 
  N.~Jarosik, C.~L.~Bennett, J.~Dunkley, B.~Gold, M.~R.~Greason, M.~Halpern, R.~S.~Hill and G.~Hinshaw {\it et al.},
  Astrophys.\ J.\ Suppl.\  {\bf 192}, 14 (2011)

\bibitem{wmap9} 
  C.~L.~Bennett, D.~Larson, J.~L.~Weiland, N.~Jarosik, G.~Hinshaw, N.~Odegard, K.~M.~Smith and R.~S.~Hill {\it et al.},
  arXiv:1212.5225 [astro-ph.CO]
  
  \bibitem{healpix}
 K.~M.~Gorski, E.~Hivon, A.~J.~Banday, B.~D.~Wandelt, F.~K.~Hansen, M.~Reinecke and M.~Bartelman,
  Astrophys.\ J.\  {\bf 622}, 759 (2005)


\bibitem{PlanckMaps} 
  P.~A.~R.~Ade {\it et al.}  [Planck Collaboration],
  arXiv:1303.5072 [astro-ph.CO].

  
\bibitem{lambda}
http://lambda.gsfc.nasa.gov/


\bibitem{PLA}
http://pla.esac.esa.int/pla/aio/planckProducts.html


\end{thebibliography}
\end{document}